\documentclass[aps,prl,showpacs,showkeys]{revtex4}
\usepackage{amsmath,amssymb,graphicx,color}

\begin{document}
\title{Revealing Hofstadter Spectrum for Graphene in a Periodic Potential}

\author{Godfrey Gumbs$^{1,2}$, Andrii Iurov$^{1}$ \footnote{{E-mail: \href{mailto:theorist.physics@gmail.com}{theorist.physics@gmail.com}}},   Danhong Huang$^{3}$ and Liubov Zhemchuzhna$^4$}
\affiliation{$^{1}$Department of Physics and Astronomy,  Hunter College  of the \\
 City University of New York, 695 Park Avenue, New York, NY 10065, USA  \\
$^{2}$ Donostia International Physics Center (DIPC),
P de Manuel Lardizabal, 4, 20018 San Sebastian, Basque Country, Spain \\
$^{3}$Air Force Research Laboratory, Space Vehicles Directorate
Kirtland Air Force Base, NM 87117, USA \\
$^{4}$Department of Math \& Physics, North Carolina Central University,
Durham, North Carolina 27707, USA}

\date{\today}

\begin{abstract}
We calculate the energy bands for graphene monolayers when electrons move
through a periodic electrostatic potential in the presence of a uniform
perpendicular magnetic field.  We clearly demonstrate the quantum fractal
nature of the energy bands at reasonably low magnetic fields. We present
results for the energy bands   as functions of both wave number and magnetic
flux through the unit cells of the resulting moir\'e superlattice. The
effects due to  pseudo-spin coupling and Landau orbit mixing by a strong
scattering potential have been exhibited.  At low magnetic fields when the
Landau orbits are much larger than the period of the modulation, the Landau levels are only
slightly broadened. This feature is also observed at extremely high magnetic fields. The density 
of states has been calculated and shows a remarkable self-similarity like the energy bands. 
We estimate that for modulation period of $10 \, nm$ the region where the Hofstadter butterfly is revealed at $B \leq 2\,T$. 
\end{abstract}



\maketitle

In recent experiments \cite{science,nature1,nature2},  graphene flake
and bilayer graphene were coupled to a
rotationally-aligned hexagonal boron nitride substrate.
The spatially varying interlayer  electrostatic potential gives
rise  to  local symmetry breaking of the carbon sublattice
as well as a long-range moir\'e superlattice potential in the graphene
layer. At  high magnetic fields, integer conductance plateaus which were
obtained at non-integer filling factors were believed to be due to
 the formation of the Hofstadter
 butterfly in a symmetry-broken Landau level.
These experiments were partially motivated by the pioneering  theoretical work
of Azbel \cite{azbel} and Hofstadter \cite{hofstadter} on
the single-particle spectrum of a two-dimensional structure in the presence of both
a periodic potential and a uniform   ambient  perpendicular magnetic field.
In this case, the energies exhibit a self-similar recursive energy spectrum. 
There are several other effects due to  the presence of boron nitride substrate.
The band gaps, which appear in graphene due to the substrate, were theoretically modeled
in \cite{McD}. The existence of a  commensurate state, when the crystal is adjusted
by the presence of the external periodic potential, has been investigated in \cite{BN4}.
Hofstadter fractal structures may also be    realized in a variety of systems, such as
Jaynes-Cummings-Hubbard lattices \cite{Hey}. We should also mention a recent study of
the energy spectrum of   Schr\"odinger electrons, subjected to general periodic potential
and   magnetic field \cite{Jan}.

\par
The  fractal nature   of the Hofstadter butterfly had captivated
researchers for many years \cite{GG,wannier,rauh,rauh2,rauh3,kohmoto1,kohmoto2,pfannkuche,
pfannkuche2,thouless,claro,schellnhuber,perschel,wu,silbernauer,sokoloff,cui1}. The  paper by
Hofstadter \cite{hofstadter} was for the energy spectrum of a periodic square lattice in the
tight-binding approximation and subject to a perpendicular magnetic field. Ever since that time,
there have been complementary calculations for the hexagonal lattice \cite{GG}, the two-dimensional
electron gas (2DEG) with an electrostatic periodic modulation potential \cite{pfannkuche,
pfannkuche2,cui1}  and even bilayer graphene where different  stacking of the two types of atoms forming the
sublattices  was considered \cite{bilayer}.  It has been claimed that one may be able to  observe evidence
of the existence of Hofstadter's butterfly in such experimentally measured quantities as density-of-states
and conductivity of the 2DEG \cite{nature1,nature2,ye}.
\par
The challenge facing experimentalists had been to carry out experiments on 2D structures
at achievable magnetic fields where the Hofstadter  butterfly  spectrum is predicted.
However, both monolayer and bilayer graphene coupled to hexagonal boron nitride
provide a nearly ideal-sized periodic modulation,  enabling unprecedented experimental
access to the fractal spectrum \cite{science,nature1,nature2}.
\par
In formulating a theoretical framework for the  energy band structure for a periodically
modulated energy band structure in a uniform magnetic field, one may adopt the procedure
of Hofstadter by using Harper's  equation which may be viewed as a tight-binding approximation
of the Schr{\"o}dinger equation.Additionally, assuming that the magnetic flux through unit cell
of the periodic lattice is a rational fraction $p/q$ of the flux quantum in conjunction with the
Bloch condition for the  wave function, one obtains a $p\times p$ Hamiltonian matrix to determine
the energy eigenvalues since one only  needs to solve the problem in a unit cell. Hofstadter himself
was concerned  about ever reaching magnetic fields where the rich self-similar structure of the butterfly
would be experimentally observed due to the estimated high fields required to achieve this.
\par
\medskip
\par
In this paper, we  supplement the recent experimental work on graphene by
first presenting a formalism for calculating  the energy band structure when
an electrostatic modulation potential is applied to a flat sheet
in the presence of a reasonably low perpendicular magnetic field. These results
are then employed in a calculation of the density-of-states (DoS).  Our results may
be verified experimentally since the DoS is directly proportional to the quantum
capacitance \cite{DOS}. The DoS may also be obtained from magnetic
susceptibility measurements. For a review of related energy band structure studies, see
\cite{sokoloff}. We also compare our results with those for a modulated
two-dimensional electron gas and discuss the difference.

 \begin{figure}
\centering
\includegraphics[width=0.85\textwidth]{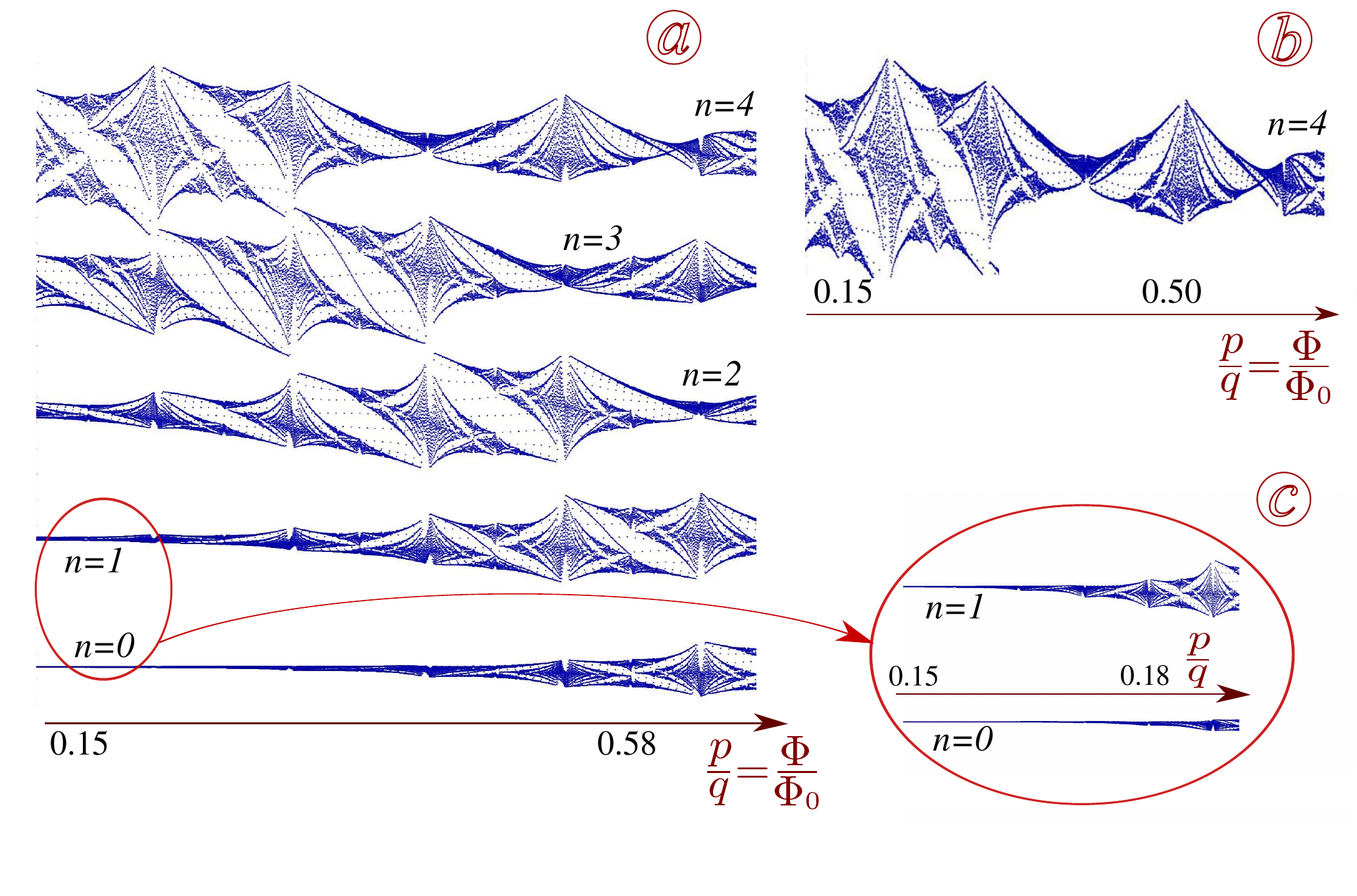}
\caption{(Color online) Energy band structure of a  weakly modulated 2DEG as a
function of magnetic    flux ratio $p/q$. Plot (a) presents the four lowest
Landau subbands for chosen modulation strength $\mathbb{V}_0=0.5 \, \hbar \omega_c$, $N=10$ and
$k_x=k_y=0.3$ in units of $2\pi/d_x$. Plot (b) shows the detailed band structure of the
$n=4$ Landau level for a 2DEG. Panel (c) shows a zoom-in of the    low-field portion
 of the two lowest levels for a 2DEG, demonstrating   self-repeated structures for all
 levels and magnetic field. }
\label{FIG:2}
\end{figure}
In the presence of a uniform perpendicular magnetic field
$B_0$ and periodic two-dimensional electrostatic  \textit{modulation potential}
defined by\,\cite{cui1}
\begin{equation}
\mathbb{V}(x,\,y)=\mathbb{V}_0\left[\cos\left(\frac{\pi x}{d_x}\right)
\cos\left(\frac{\pi y}{d_y}\right)\right]^{2N}\ ,
\label{E1}
\end{equation}
where $N=1,\,2,\,\cdots$ is an integer determining the size of the scatterers, the parameter
$\mathbb{V}_0$ is the modulation amplitude, and $d_x,\,d_y$ are the modulation periods in the $x$
and $y$ directions, respectively, we rewrite the Hamiltonian operator as
\begin{equation}
\mathcal{H} =
\left[
\begin{array}{cc}
\mathbb{V}(x,y) & \hat{p}_x+eB_0y\hat{x}_0 + \imath \hat{p}_y \\
\hat{p}_x+eB_0y\hat{x}_0 - \imath \hat{p}_y & \mathbb{V}(x,y)
\end{array}
\right]
\end{equation}
For this new system, the magnetic flux per unit cell is $\Phi=B_0(d_x d_y)$, which is assumed
to be a rational fraction of the flux quantum $\Phi_0=\hbar/e$, i.e.,
$\beta\equiv\Phi/\Phi_0=p/q$, where $p$ and $q$ are prime integers. Furthermore, we choose
the first Brillouin zone defined by $|k_x|\leq \pi/d_x$ and $|k_y|\leq \pi/(qd_y)$.
\par
\medskip
By using
the Bloch-Peierls condition, the wave function of this system may be expanded as
\begin{equation}
\Phi^{\pm}_{\ell;\,n,{\vec k}_{||}} \left( {x,\,y} \right)=\frac{1}{\sqrt{2{\cal N}_y}}
\sum\limits_{s=-\infty}^{\infty} \texttt{e}^{ik_y\ell_B^2(sp+\ell)K_1}\Psi_{n,k_x-(sp+\ell)K_1}^{K,\pm} \left( {x,\,y} \right)
\end{equation}
where ${\vec k}_{||}=(k_x,\,k_y)$, ${\cal N}_y=L_y/(qd_y)$ is the number of unit cells,
which are spanned by $b_1=(d_x,\,0)$ and $b_2=(0,\,qd_y)$, in the $y$ direction,
$L_y$ ($\to\infty$) is the sample length in the $y$ direction,
$K_1=2\pi/d_x$ is the reciprocal lattice vector in the $x$ direction and
$\ell=1,\,2,\,\cdots,\,p$ is a new quantum number for labeling split $p$ subbands
from a $k_x$-degenerated landau level in the absence of modulation.
The above wave function satisfies the usual Bloch condition:
\begin{equation}
\Phi^{\pm}_{\ell;\,n,{\vec k}_{||}} \left( {x+d_x,\,y+qd_y} \right)={\rm e}^{ik_xd_x}\,{\rm e}^{ik_yqd_y}\,\Phi^{\pm}_{\ell;\,n,k_{||}} \left( {x,\,y}\right) \, .
\label{Bloch}
\end{equation}

Since the wave functions at $K$ and $K^\prime$ points are decoupled from each other
for monolayer  graphene, which is different from   bilayer graphene,\,\cite{petters}
we can write out explicitly the full expression for the wave function at these two points.
A tedious but straightforward calculation yields
the magnetic band structure for this modulated system as a solution of the eigenvector problem
${\cal M} \bigotimes\vec{\cal A}(\vec{k}_{||})=0$ with the coefficient matrix $\tensor{\cal M}$
given by

\begin{equation}
\{{\cal M}\}_{j,\,j'}=\left[E^{\mu}_n-\varepsilon({\vec k}_{||})\right]
\delta_{n,n'}\delta_{\ell,\ell'}\delta^{(n)}_{\mu,\mu'} +\
  \mathbb{V}^{\ell',n',\mu'}_{\ell,n,\mu}(\vec{k}_{||})\ ,
\label{MM}
\end{equation}
where $\delta^{(n)}_{\mu,\mu'}=1$ for $n=0$ and $\delta^{(n)}_{\mu,\mu'}=\delta_{\mu,\mu'}$ for $n>0$,
 $j=\{n,\,\ell,\,\mu\}$ is the composite index, and
$\{\vec{\cal A}(\vec{k}_{||})\}_j={\cal A}^\mu_{n,\ell}(\vec{k}_{||})$  
is the eigenvector.  The eigenvalue\
$\varepsilon_\nu({\vec k}_{||})$ of the system is determined by ${\rm Det}\{\tensor{\cal M}\}=0$. Here we
calculate $\mathbb{V}^{\ell',n',\mu'}_{\ell,n,\mu}(\vec{k}_{||})$ as the Fourier transform. 
The detailed derivation
of these matrix elements could be found in \cite{our1}. 
 \begin{figure}
\centering
\includegraphics[width=0.85\textwidth]{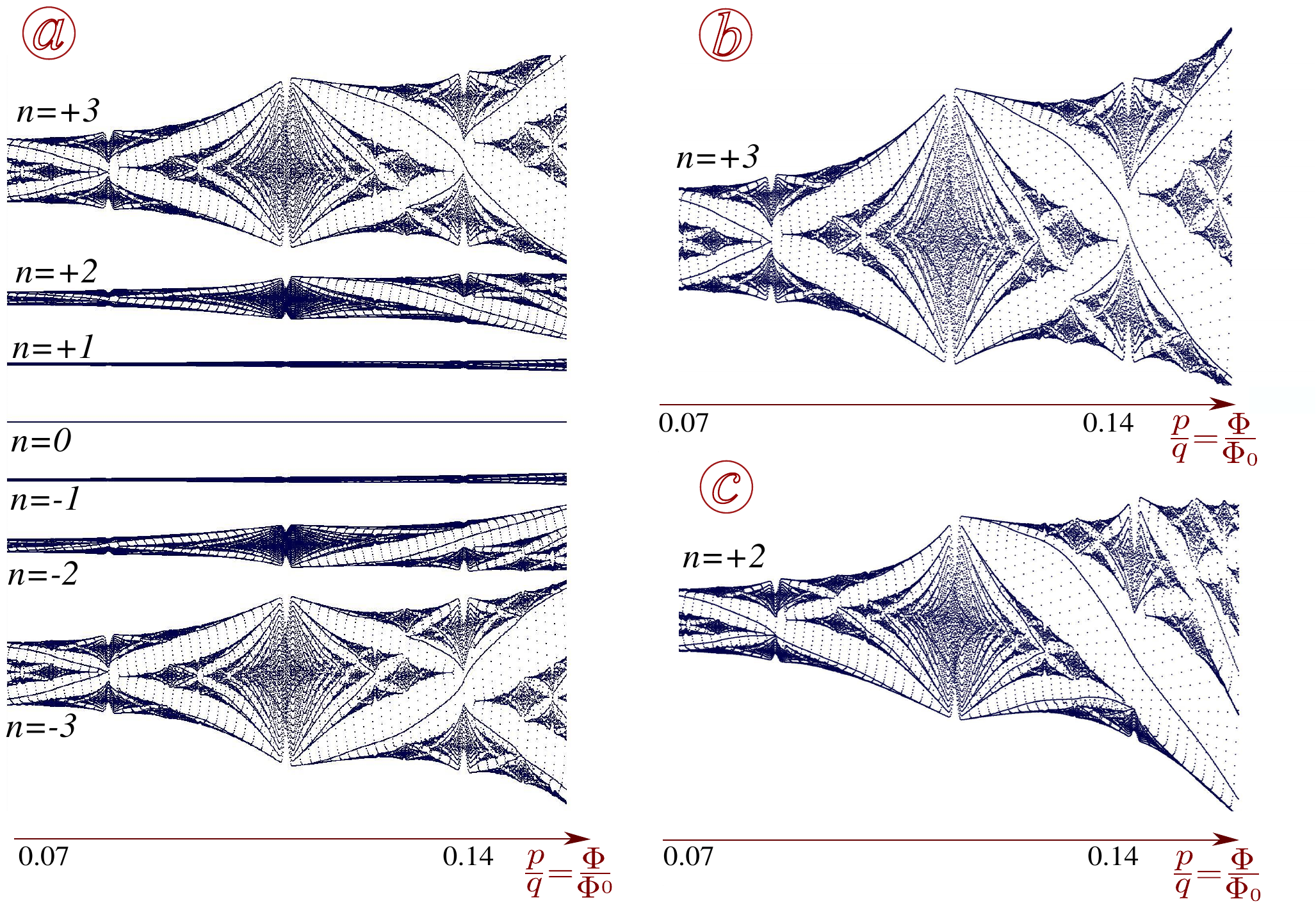}
\caption{(Color online) Band structure of  strongly modulated graphene as a function
 of magnetic  flux ratio $p/q$. Plot (a) gives the lowest few Landau subbands
 from both valence and conduction bands for a chosen modulation $\mathbb{V}_0=2.0 \, \hbar \omega_c$
  and $k_x=k_y=0.3/d_x$. Plots (b) and (c) show details of the band structure of
  the $n=+2$ and $n=+3$  Landau levels. The levels are not mixed.}
\label{FIG:3}
\end{figure}

 \begin{figure}
\centering
\includegraphics[width=0.6\textwidth]{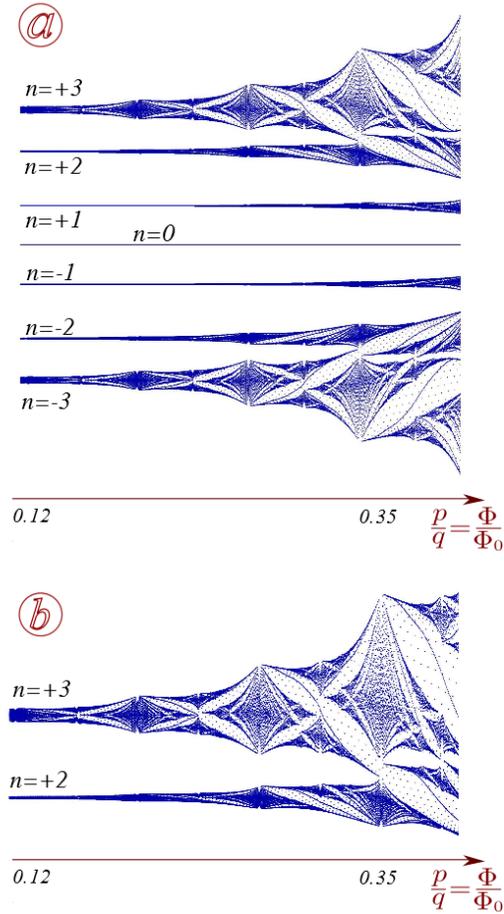}
\caption{(Color online) (Color online) Band structure of a modulated graphene monolayer
as a function of magnetic   (flux ratio $p/q$). Plot (a) demonstrates the lowest
few Landau subbands from both valnece and conduction bands for a chosen modulation
$\mathbb{V}_0=5.0 \, \hbar \omega_c$ and $k_x=k_y=0.3$. Plot (b) and (c) show the structure
of $(n=+2)$ and $(n=+3)$ Landau level. The levels are mixed.}
\label{FIG:4}
\end{figure}

Fig. \ref{FIG:2} shows that  for a 2DEG, the lowest perturbed Landau subband
which originates from the unperturbed $n=0$ Landau level
merges with the resulting butterfly spectrum at the  highest magnetic field compared to
the $n=1,2,3,4$ Landau levels in the conduction band.  The onset of the butterfly takes  place
around $p/q=1/5$ which would correspond to  a magnetic field $B\approx 2$ T for $d_x=d_y=10$ nm.
Furthermore, our calculations have shown that for graphene the symmetry between the valence and conduction
bands is destroyed by modulation. There is always mixing of the subbands regardless of the value for $\mathbb{V}_0$.
This is in contrast with modulated 2DEG where for weak $\mathbb{V}_0$, the Landau subbands do not overlap
as shown in Fig.\ \ref{FIG:3}. The lowest subband is shifted upward like the other
subbands but is not widened as much as the higher subbands. The feature of self-similarity is also
apparent in the excited subbands at intermediate magnetic fields. There is only a shift and broadening
of the subbands in the low and high magnetic field regimes for modulated 2DEG.

\begin{figure}
\centering
\includegraphics[width=0.49\textwidth]{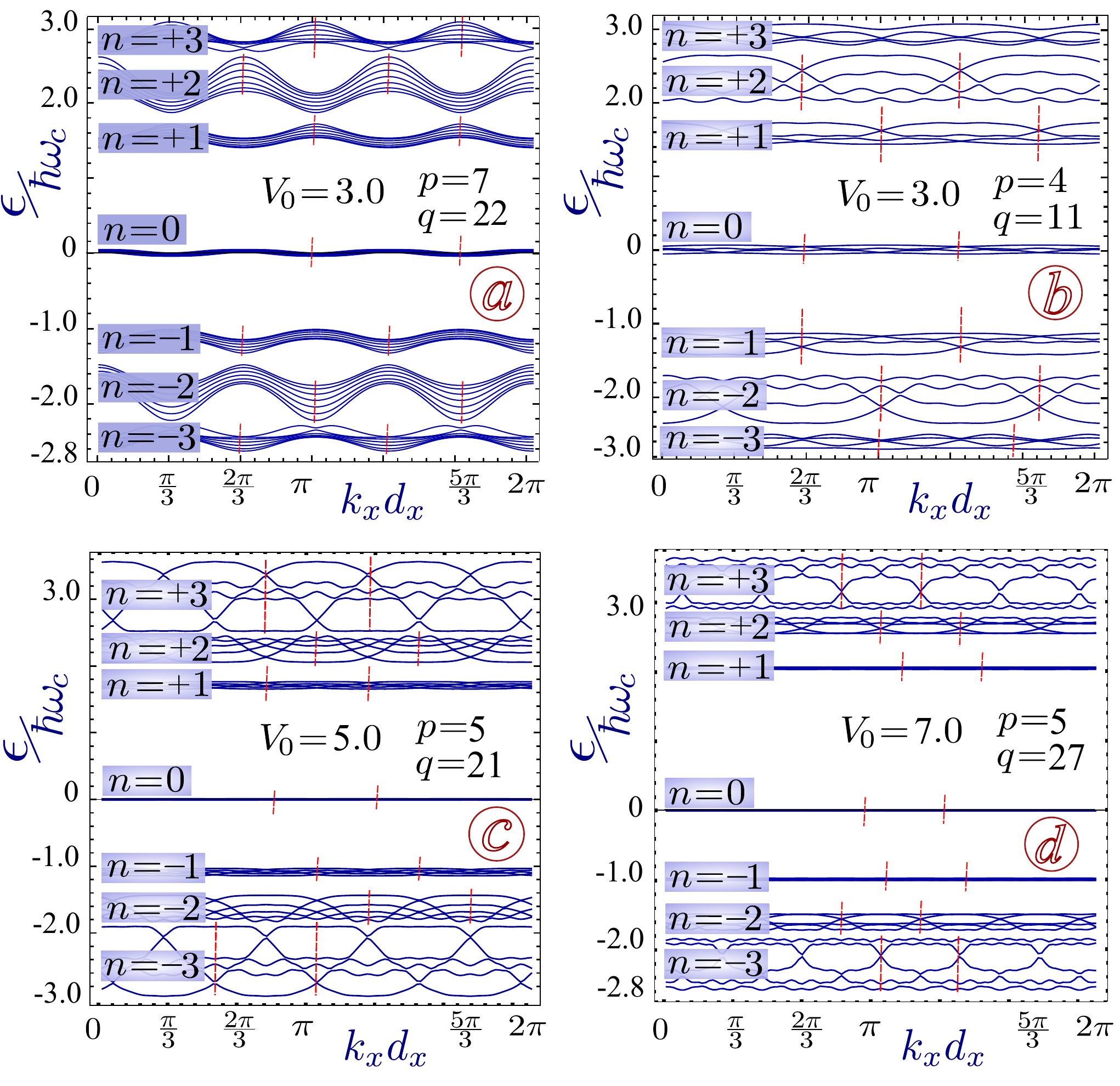}
\caption{(Color online) Energy dispersion as functions of $k_x d_x$ for graphene
and 2DEG with chosen values of modulation potential $\mathbb{V}_0$ and magnetic flux $p/q$
in units of the flux quantum. The energy is scaled in terms of $v_F\sqrt{B e \hbar}$.}
\label{FIG:5}
\end{figure}

The results of our calculations for  the energy eigenvalues  of
modulated graphene  as a function of magnetic flux appear in Fig. \ref{FIG:3}. We included  the $n=0,\pm 1,\pm2,
\pm3,\pm 4$ as we did in obtaining Fig.\ \ref{FIG:2}. For weak magnetic fields, the Landau
levels in both valence and conduction bands  are  slightly broadened into narrow  subbands
but shifted upward by the perturbing potential $\mathbb{V}_0$. Another effect due to modulation
is to cause these Landau bands to have negative slope at weak magnetic fields
which then broaden enough at higher magnetic fields to produce Landau orbit mixing,
reflecting the  commensurability   the magnetic and lattice Brillouin zones.
In Fig.\ \ref{FIG:4}, we demonstrate a Hofstadter dispersion plot for the case when
at least two Landau levels are mixed, featuring a larger fractal self-repeated structure,
which incorporates more than one energy level. 

\par
In Fig.\ \ref{FIG:5}, we present the dispersion curves as  a function of $k_x d_x$
for  chosen value of $\mathbb{V}_0$ and two pairs of values of $p$ and $q$  corresponding to two
different magnetic field strengths.  In each case,  there are $p$ Landau subbands,
$q/p$  determines the number of oscillation periods in the first  Brillouin zone
for each of these subbands. Both the valence and  conduction subbands are shifted upward
but the conduction subbands are shifted more than the valence subbands for each corresponding
Landau label for the unmodulated structure.  This shift is increased when  the modulation amplitude
is increased. The original zero-energy Landau level is  only slightly broadened
and is the least affected by $\mathbb{V}_0$. If the sign of the modulation amplitude is
reversed to correspond to an array of quantum dots, then the subbands are  all shifted downward.
from their positions for an unmodulated monolayer graphene.
\par
\medskip
\par
\begin{figure}
\centering
\includegraphics[width=0.5\textwidth]{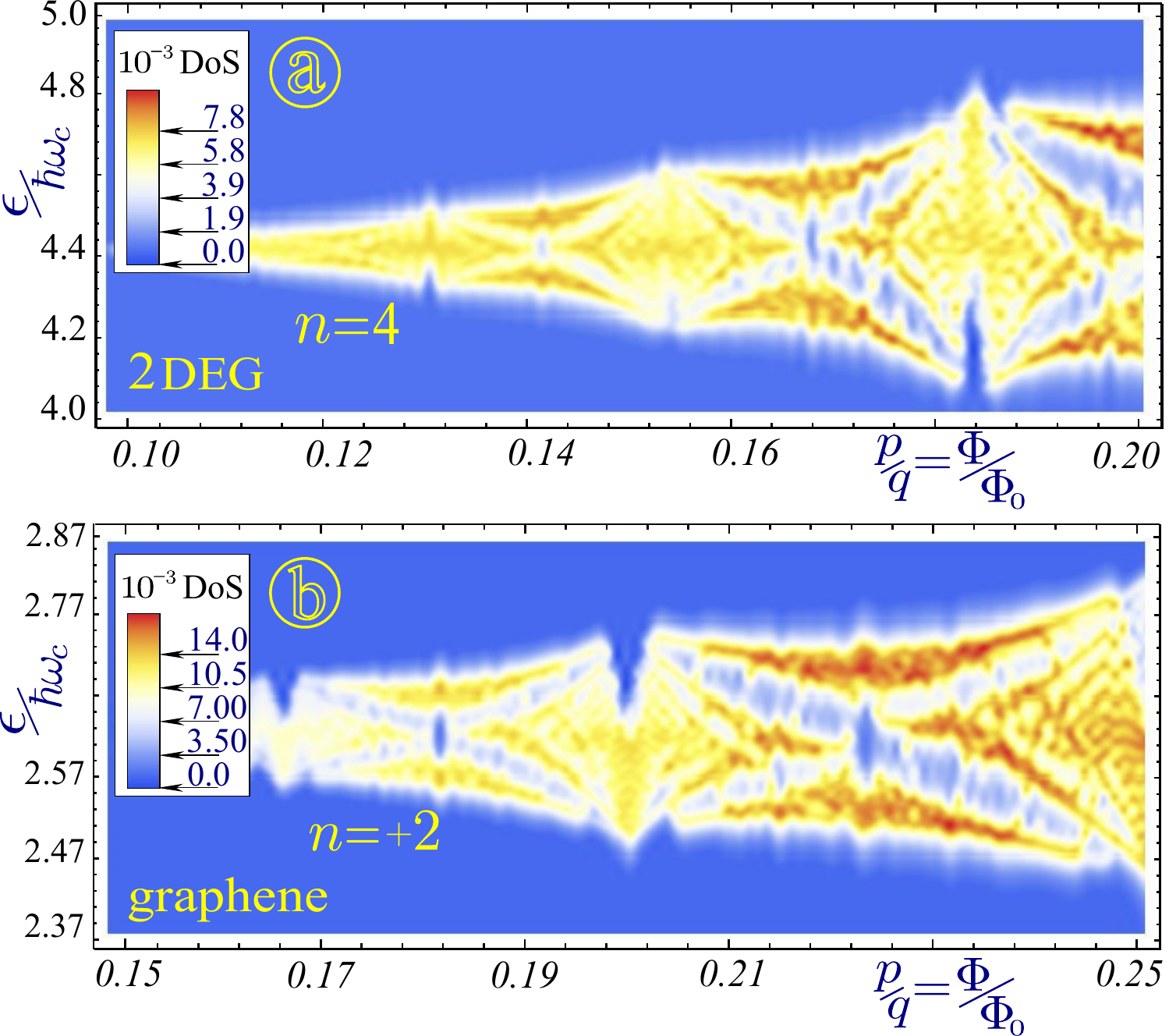}
\caption{(Color online) (Color online) Density-of-states plots for modulated
2DEG and graphene. Panel (a) shows the density of states for a 2DEG  with modulation
$\mathbb{V}_0=0.5 \, \hbar \omega_c$ and   modulation parameter $N=5$. Plot (b) demonstrates
the corresponding situation for graphene with $\mathbb{V}_0=1.5 \, \hbar \omega_c$ and $N=3$.
For both plots (a) and (b), the density-of-states amplitude is increased proportionally
 to the value of magnetic field in order to make the Hofstadter structure visible at
 large fields.}
\label{FIG:6}
\end{figure}

We also calculate and display the density-of-states plots, which demonstrate the general
magnetic field dependence of the Hofstadter spectrum, showing the general fractal structures
independent of the specific values of the wave vector $\vec{k}_{\parallel}$.
Based on the calculated eigenenergy $\varepsilon_\nu({\vec k}_{||})$, we can further
calculate the electron density of states in this system, given by

\begin{equation}
\rho(\varepsilon,\,B_0)=\frac{1}{2\pi^2qd_xd_y}\,\sum\limits_{\nu}\int\limits_{-\pi/d_x}^
{\pi/d_x} dk_x\int\limits_{-\pi/(qd_y)}^{\pi/(qd_y)}dk_y\,\frac{\Gamma/\pi}{[\varepsilon-
\varepsilon_\nu({\vec k}_{||})]^2+\Gamma^2}\ ,
\end{equation}
where $\Gamma$ represents the level broadening.
\par
Making use of  our calculated energy eigenvalues $\epsilon_{\nu}(\vec{k} )$,
we further determine  the electron   density-of-states  for monolayer
graphene in the presence of a uniform perpendicular magnetic field.
\par
\medskip
\par
A plot of the density-of-states as a function of magnetic field basically
reproduces the Hofstadter self-repeating fractal structure, averaged over all
allowed values of $\vec{k} $. In our calculations, the delta
function is chosen as a Lorentzian.   This  leads to the finite width for  the
density of states for various values of energy $\epsilon$.  In this
regard,  one should look at Ref. \cite{DoSNemec}, in which the density-of-states has been 
calculated 
for carbon nanotubes for the various cases of magnetic field strength and orientation.

\par
Fig. \ref{FIG:6}  shows $\rho(\epsilon, B)\, B$. We chose this specific
set of results obtained for the density-of-states to demonstrate  the effect
 due to the  modulation in order to show how  the Hofstadter
structure may be  suppressed by the strong $\delta-$like peaks at low magnetic fields.
 \par
\medskip
\par
 
 Finalizing our description of the impact of the electrostatic modulation on the 
 Hofstadter spectrum, we would like to comment on how the standard Dirac cone type of 
 energy dispersion would be modified in the presence of modulation
only, without the magnetic field. This situation could be described by the following 
Hamiltonian: $\mathcal{H}_{\mathbb{V}} = \sigma \cdot {\bf p} + 
\mathbb{V}(x,y) \mathcal{I}$ with $\mathbb{V}(x,y)$ given in Eq.\eqref{E1}
and $\mathcal{I}$ is the unit matrix. Obviously, we are dealing with the continuous 
spectrum, accompanied generally by a strong anisotropic dependence on $k_x$ and $k_y$ 
wave numbers. The wave function, corresponding to such periodic potential also satisfied
the Bloch condition \eqref{Bloch}.
 Such a Hamiltonian with a periodic potential in one dimension was considered in  
 \cite{FertigMain}. Their results show that new zero energy states emerge and the 
 wave function corresponds to an overdamped particle in a periodic potential.
These zero energy solutions in fact represent new Dirac points. The presence of the 
zero-averaged wavenumber gaps as well as extra Dirac points in the band structure 
for the graphene-based one-dimensional superlattices were also found in \cite{minigaps}.  
The effect of non-homogeneous magnetic and electric fields was addressed in \cite{fields2}.

\par
\medskip
\par
In summary,  the well established Dirac fermion model is utilized
to investigate the Landau level spectra of monolayer graphene in
the presence of a periodic electrostatic potential. The intrinsic
pseudospins from different Landau orbits which mix effectively give
rise to multiple splitting of Landau levels. By incorporating Bloch
wave function characteristics, we  established an eigenvalue equation
which yields fractal self-similar structure for the allowed energy
band structure determined by the orbital pseudo spin and magnetic field
signatures. In our calculations for the density-of-states, the physical
origins of  self-similarity  are clearly established as being accessible
experimentally. In particular, the emergence of Hofstadter's butterfly
spectrum lies within a reasonable range of magnetic field that is currently
available.  Our numerical results clearly demonstrate magnetic field control
of the energy density locations of the charge carriers and provide a basis
for future experiments where  regions of high absorption and conductivity may
be observed at certain field strength. On the contrary, in the absence of magnetic
field, the density-of-state lines are aligned next to each other.

\acknowledgments
This research was supported by  contract \# FA 9453-13-1-0291 of
AFRL.

\end{document}